\documentclass[prb,twocolumn,showpacs,preprintnumbers,amsmath,amssymb]{revtex4-1}
\usepackage{graphicx}	
\usepackage{bm}	
\usepackage{amssymb}
\usepackage{multirow}
\usepackage{array}
\usepackage{dcolumn}	
\usepackage[font=small,skip=-5pt]{caption}

\begin{document}

\title{Experimental Determination of the Critical Spin Glass Correlation Length
\\ in Single Crystal CuMn}

\author{G. G. Kenning$^1$}
\author{D. L. Schlagel$^2$}
\author{V. Thompson$^1$}
\affiliation{$^{1}$Department of Physics,~Indiana University of Pennsylvania, Indiana, Pennsylvania, 15705}
\affiliation{$^2$Division of Materials Science and Engineering, Ames Laboratory, Ames, IA 50011}

\date{\today}   

\begin{abstract}
	
\noindent

Using high resolution SQUID magnetometry we have made detailed measurements of the waiting time effect of the  Thermoremanent Magnetization (TRM) decays on a single crystal CuMn(6$\%$) spin glass sample near $T_g$.  We have systematically mapped the rapid decrease of the characteristic timescale $tw_{eff}$, approaching $T_g$ from below, for waiting times ranging from 100 s to 100000 s. Using $tw_{eff}$ to determine the length scale of the growth of correlations during the waiting time, $\xi_{TRM}$, (observed in both numerical studies and experiment), we observe both growth of $\xi_{TRM}$ in the spin glass phase and then a rapid reduction very close to $T_g$. We interpret this reduction in $\xi_{TRM}$, for all waiting times, as being governed by the critical correlation length scale $\xi_{crit}=a(T-T_c)^{-\nu}$.
	
\end{abstract}


\maketitle
\noindent {\bf I.~Introduction}\\

In 1972 Cannella and Mydosh \cite{Cannella72} found evidence of a phase transition (through observation of a cusp in the magnetic susceptibility) in dilute random magnetic alloys of Au with a few percent of Fe. Further investigations found this cusp to be time dependent.\cite{Mulder81}  In 1976 Edwards and Anderson\cite{Edwards75} (EA) showed that a magnetic system with random couplings can have a phase transition. Following these seminal contributions, many different  experiments were performed on spin glasses\cite{Binder86} and the system proved to be experimentally interesting due to  a number of time-dependent effects spanning the complete range of experimental timescales available. In contrast to the susceptibility measurements, specific heat measurements displayed no evidence of a phase transition (expected from EA) leading to confusion about the nature of the transition. Also missing from the picture are  measurements of a critical correlation length $\xi_{crit}(T)$. This is likely due to the lack of a probe that can couple to the effectively random spin configurations (paramagnetic like) that exist within the spin glass state. 

In 1983 two groups\cite{Cham83, Lund83} observed that the Thermoremenant Magnetization (TRM) exhibited a decay  that was dependent on the length of time the sample was held at the measuring temperature before the  magnetic field was turned off. This is known as the waiting time effect. To measure the TRM, the material is  cooled along the Field Cooled Magnetization $M_{FC}$ line (in a magnetic field H)  though a temperature $T_g$ (the onset of irreversible behavior), to a temperature T and held for a waiting time tw. In small magnetic fields, the magnetization ($M_{FC}$) is approximately constant below $T_g$, indicating that all of the spins are effectively frozen.  The magnetic field is  shut off and the magnetization decay recorded. Within the spin glass phase (T $<$ .9$T_g$), the time dependencies of this effect appeared to be independent of temperature, a large departure from Arrhenius behavior often seen in materials dynamics. This observation has lead to the development of the TRM as a powerful probe of the spin glass state, elucidating such questions as the structure of energy barriers \cite{Led91}, the nature of aging in the spin glass phase\cite{Alba87}, the effect of magnetic field on time dependencies\cite{Joh99}, memory\cite{Dupuis01} and rejuvenation\cite{Lefloch94} effects, and the development of the $S(t)=-dM/dln(t)$\cite{Nord86} function as a probe of time dependent effects. The S(t) function is a  straightforward method of observing the waiting time effect.   The S(t) function displays a peak at a time equal to the time where an inflection point in the decay is observed.  In the temperature range 0.4-0.9$T_g$, this characteristic time scale is observed to occur at a time approximately equal to the input waiting time. In this paper we use the S(t) function to investigate time and spatial  dependencies in the spin glass state near $T_g$, in particular focusing on the region T $ > $ 0.9$T_g$. 

In 1996 Kisker et al.\cite{Kisker96} analyzed 2D and 3D numerical simulations of Ising spin glass models. They found that they  could determine a spatially dependent correlation length scale
using a 4-spin autocorrelation function,
 \begin{equation}
G_T(r,t_w)= \frac{1}{N}\displaystyle{\sum_{i=1}^N\frac{1}{t_w}\sum_{t=t_w}^{2t_w-1}\big[\langle S_i^a(t)S_{i+r}^a(t)S_i^b(t)S_{i+r}^b(t)\rangle \big]_{av}}.
\end{equation}
These spatial correlations grow according to 
\begin{equation}
\xi(t_w, T)=A\left({\frac {t_w}{\tau_0}}\right)^{c_2(T/T_c)}~~,
\end{equation}
where $\tau_o$ is a microscopic exchange time, and $A$ and $c_2$ are  constants. 

In the confining geometry of a 15.7 nm Ge:Mn  thin film, Guchait and Orbach\cite{Guchhait14} found that for waiting times larger than a crossover time  $t_{co}$  the ZFC decay became exponential (indicating Arrhenius behavior and the existence of a maximum energy barrier which governed the decay). Setting the limiting length scale for the growth of correlations Eq.(2) to $\xi_{ZFC}(t_{co}, T) = \mathcal{L}$ the film thickness, they determined the maximum observed energy barrier  $\Delta_{max}$ from Joh et al. \cite{Joh99}

\begin{equation}
\displaystyle{\frac{\Delta_{max}(\mathcal{L})}{k_B T_g}} = \frac{1}{c_2}\Bigg[\ln\bigg({\frac{\mathcal{L}}{a_0}}\bigg) - \ln{c_1}\Bigg],
\end{equation}
where  the constant A, in Eq. 2, has been replaced by $c_1a_o$ and   $\xi= \xi_{ZFC}$.  $a_o$  is the average distance between magnetic ions (allowing for comparison of data taken  at different concentrations hence different $T_g$.  This dynamic analysis was extended to CuMn(14$\%$) thin films by Zhai et al.\cite{Zhai17} who found consistent results for three films with substantially different $\mathcal{L}$, using c$_1$= 1.448 and c$_2$ = .104. They also associated the maximum barrier  with the observed thin film freezing temperatures $\Delta_{max} = k_bT_f(\mathcal{L})$ from Ref[16] and found that Eq. 3 substantially predicts the form of $T_f(\mathcal{L})$. In the above studies it is assumed that $\xi$ grows isotropically until it reaches the thickness of the film.  For the rest of this manuscript, superposition\cite{Mathieu01} is assumed and $\xi = \xi_{TRM} =\xi_{ZFC}$.

In a previous study,\cite{Kenning18} we found that the time associated with the peak in the S(t) function ($tw_{eff}$) dramatically decreased above 0.9$T_g$. It was conjectured that the correlation length may be reaching the polycrystalline size scale and the dramatic decrease may be due to finite size effects.\cite{Tennant20} To test this hypothesis we grew and prepared a single crystal CuMn(6$\%$) sample. The sample was prepared using the Bridgman method. The Cu and Mn were arc melted several
times in an Argon environment and cast in a copper mold.
The ingot was then processed in a Bridgman furnace.
XRF (X-ray fluorescence) and optical observation
showed that the beginning of the growth is a single phase. Further details on the production of the sample are presented elsewhere.\cite{Zhai19}

In this study we differentiate between the time tw, the experimental time spent in a field of 5G before setting the field to 0G and $tw_{eff}$ which is determined by the peak in the S(t) function. We also differentiate between $T_g$ ($T_g$=31.5 K for the single crystal sample as determined from FC/ZFC measurements\cite{Zhaiprivate}), the temperature where remanent behavior is first observed, and $T_c$ the critical phase transition temperature. 

\bigskip
\maketitle
\noindent{\bf II. Experimental Results }\\
\begin{figure}[t] 
\resizebox{8.6cm}{6cm}{\includegraphics{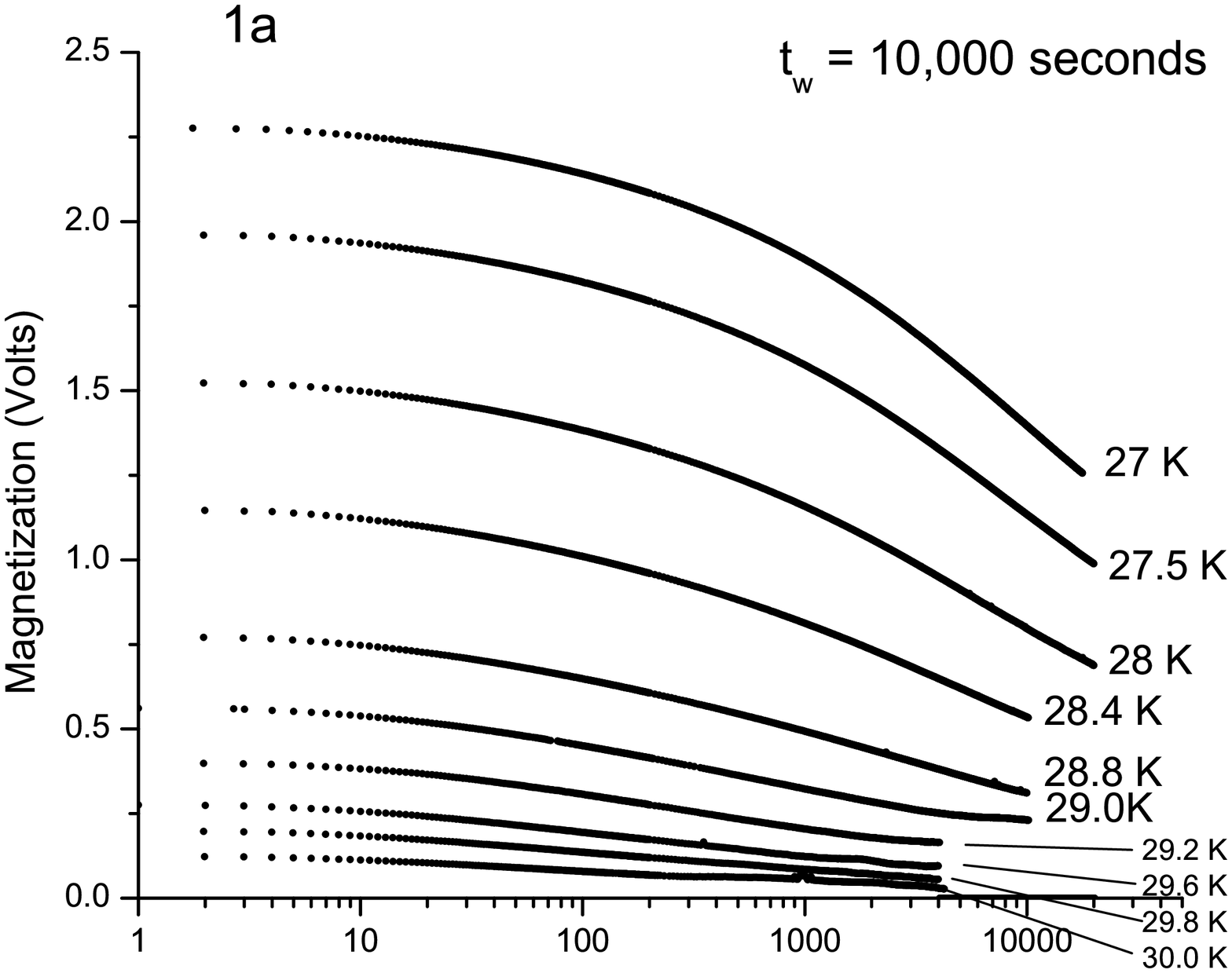}}
\resizebox{8.6cm}{6cm}{\includegraphics{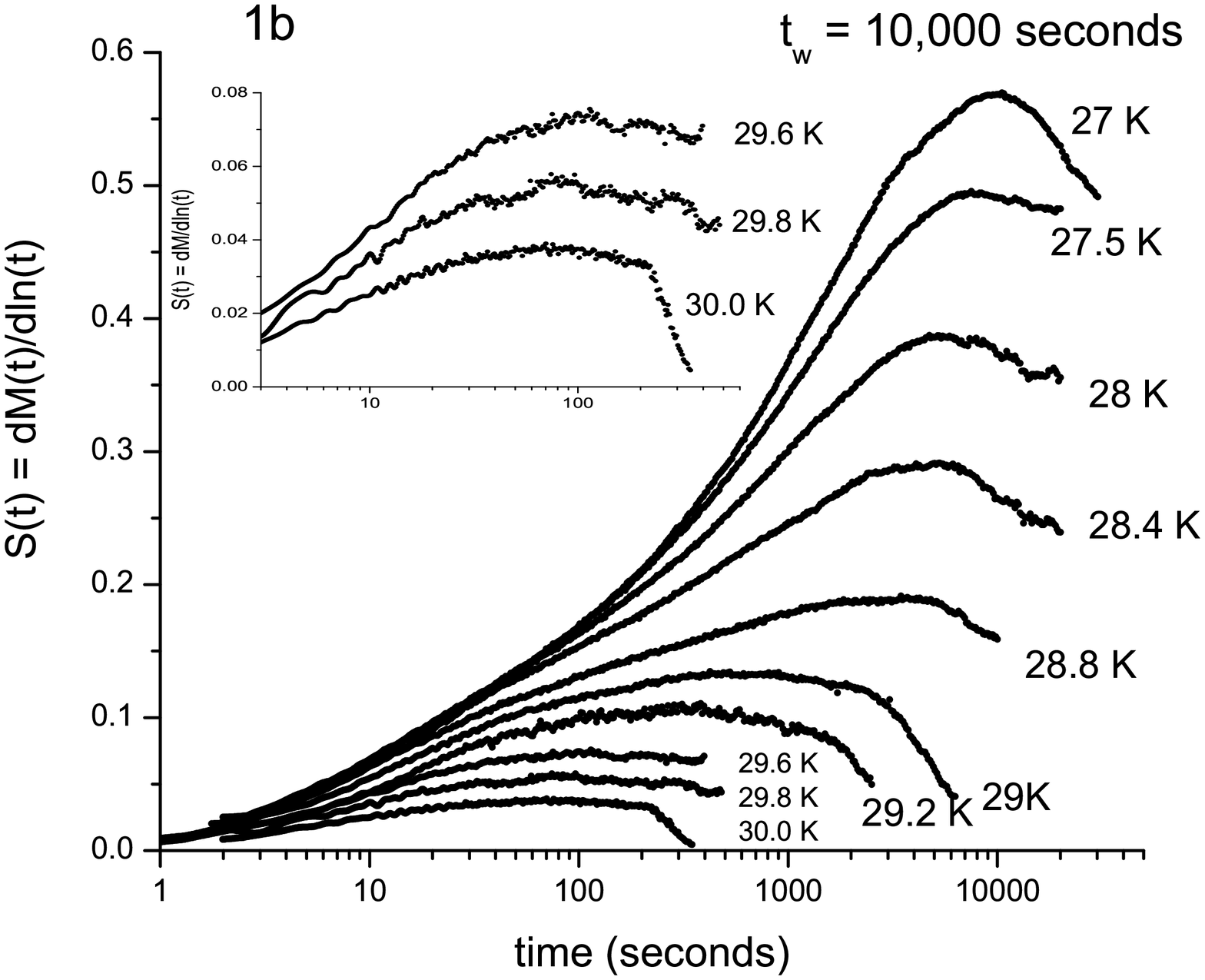}}
\smallskip
\captionsetup{format=plain,justification=centerlast} 
\caption{TRM decays at temperatures ranging between 27 K and 30 K for the 10,000 s waiting time, $T_g=31.5K$. Figure 1b displays the S(t) function of the same data. The inset in figure 1b shows the three highest temperature data sets on an expanded scale. }
\end{figure}

All data presented were taken using The Indiana University of Pennsylvania (IUP) High Sensitivity Dual DC SQUID magnetometer. The magnetometer uses Quantum Design (QD) DC SQUIDS coupled to 1 cm diameter pickup coils in a 2nd order gradiometer configuration. The magnetometer continously monitors the change in magnetic flux in the pick up coils, over the entire measurement time, and is queried once per second to record the data. The magnetization signal is output in volts. When measuring near the limits of the magnetometer's sensitivity limit, this technique is sensitive to  atmospheric pressure changes. We  can significantly reduce this unwanted signal by inserting an electronic pressure control valve downstream of the He boiloff line and pumping on the output. The measurement temperature can be controlled  for more than 100,000 s with mean fluctuations less than $\pm 1 mK$. Side by side comparisons with a commercial QD DC SQUID magnetometer located at The University of Texas indicate  that the IUP magnetometer has a signal to noise ratio approximately 27 times better that the QD system. This improved signal to noise is especially useful in this study as the signal rapidly decreases as Tg is approached. Other details of the experimental apparatus are discussed elsewhere.\cite{Kenning18}

TRM data were acquired in two separate series of experiments, both using a 5 G field. This insured that $T_g(H)$ is constant.\cite{Kenning91} The first set of data were taken  over temperatures ranging from 7 K (0.22$T_g$) to 34 K (1.1$T_g$) with probed waiting times of 100 s, 1000 s and 10,000 s and  measuring times ranging from 20,000-100,000 s. A second series of measurements were taken in  the vicinity of $T_g$ on a grid as small as 0.1 K (tw=1000 s), although most of the waiting times were probed at 0.2 K increments. Fig. 1 displays TRM decay data (Series 2) for TRM decays with waiting time tw =10,000 s for temperatures between 27K and 30K. In this temperature region both the magnetization and $tw_{eff}$ decline by several orders of magnitude.

\begin{figure}[t] 
\resizebox{9cm}{8cm}{\includegraphics{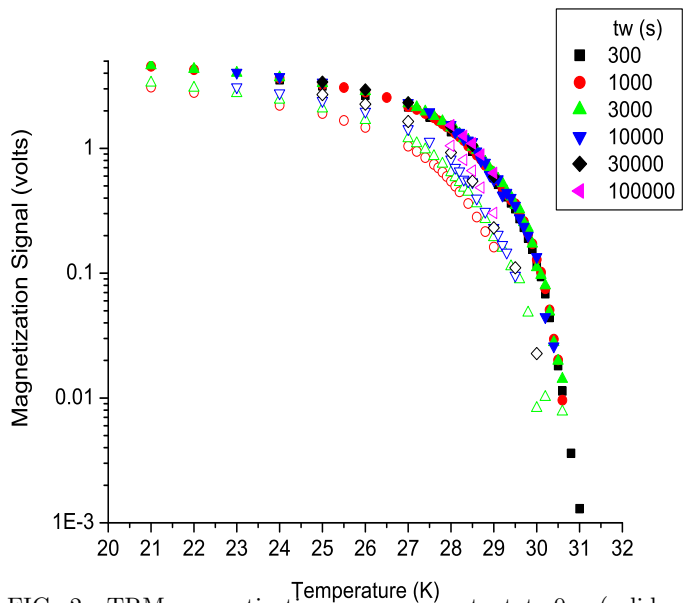}}
\captionsetup{format=plain,justification=centerlast} 
\caption[]{TRM magnetization measurement at t=0 s (solid symbols) and the TRM magnetization measurement at t=10,000 s (open symbols). The difference between these measurements display both the waiting time effect and the magnitude of the TRM decay.}
\end{figure}

The TRM decay has three distinct regions. When the field is cut off there is a large rapid decay which is  waiting time independent but strongly temperature dependent. This reversible decay\cite{Kenning95} (often called the stationary decay) is approximately the same magnitude as the  Zero Field Cooled Magnetization $M_{zfc}$ signal. At low temperatures this signal can be as small as 0.5$M_{FC}$ (at approximately 0.3$T_g$) increasing with temperature up to $\approx 0.97M_{FC}$ at 0.9$T_g$ and finally equaling $M_{FC}$ at $T_g$. The second distinct region of the decay is the waiting time dependent decay. While this term has structure near a time comparable to the waiting time, this decay component can extend out several orders of magnitude in measurement time greater than the waiting time. Finally, there is a  waiting time independent logarithmic decay that makes up the residual remanence.\cite{Kenning06}
\begin{figure}[t] 
\centering
\resizebox{9cm}{8cm}{\includegraphics{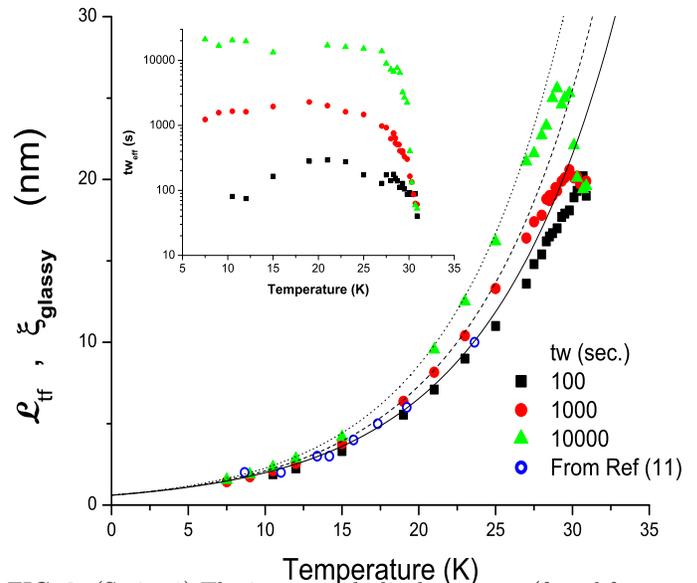}}
\captionsetup{format=plain,justification=centerlast} 
\caption[]{ (Series 1) The inset graph displays $tw_{eff}$ (found from the peak of the S(t) function), versus temperature for waiting times ranging of 100 s, 1000 s and 100,000 s. The correlation length $\xi_{TRM}(tw_{eff})$ as a function of temperature (main graph). The open circles are data from Ref[16] and plot the thin film thickness as a function of observed freezing temperature. The solid line is a best fit to the thin film data to determine $c_1$ and $c_2$. The dashed line is a fit to Eq. 2 with tw=1000 s. The dotted line plots Eq. 2 with tw=10,000 s.}     
\end{figure}

Figure 2 displays the static TRM values taken at the first point of the decay 
Mo (t = 0 s), (solid symbols), and at 10,000 s into the decay, M(t = 10,000 s) (open symbols). It can readily be observed that as Tg (31.5 K) is approached from below, the TRM magnetization decreases by several orders of magnitude. The magnitude of the TRM decay and the waiting time effect are evident in the difference between the 
t = 0 s and t = 10,000 s data points.  While the waiting time effect is effectively over for small waiting time and 10,000 s measuring time, the tw=10,000 s data is only approximately at the inflection point. The waiting time independent logarithmic decay makes up the majority of the residual remanence left after the 10,000 s measuring time.\cite{Kenning06} It can be observed that for most of the spin glass phase, the TRM decay is only a small contribution to the entire remanence.

In the inset of Fig. 3 the effective waiting times (Series 1) for  T $>$ 0.22$T_g$, are plotted as a function of temperature. At low temperatures $tw_{eff}$ (time associated with the peak in the S(t) function) is a little larger than tw but approximately constant over a wide temperature range 0.22$T_g$ $<$ T $<$ 0.9$T_g$ indicating the standard waiting time effect. Above 0.9$T_g$ we observe the same rapid decline of $tw_{eff}$ that was observed in the polycrystalline bulk sample. Since this sample is a single crystal (0.3mm x 0.5mm) this effect is clearly not due to finite size effects due to crystallites. 

The open circles in the main plot of Fig. 3 are  thin film thicknesses $\mathcal{L}$  as a function of the measured freezing temperature $T_f$, from Ref[16]. The estimated timescale of the  $T_f$ measurements was approximately 200 s.  We obtain the values $c_1$=.87 and $c_2$= .11 by fitting Eq. 2 with tw=200 s (solid line), to the thin film data. These values are used for the following analysis.
\begin{figure}[t] 
\centering
\resizebox{8cm}{8cm}{\includegraphics{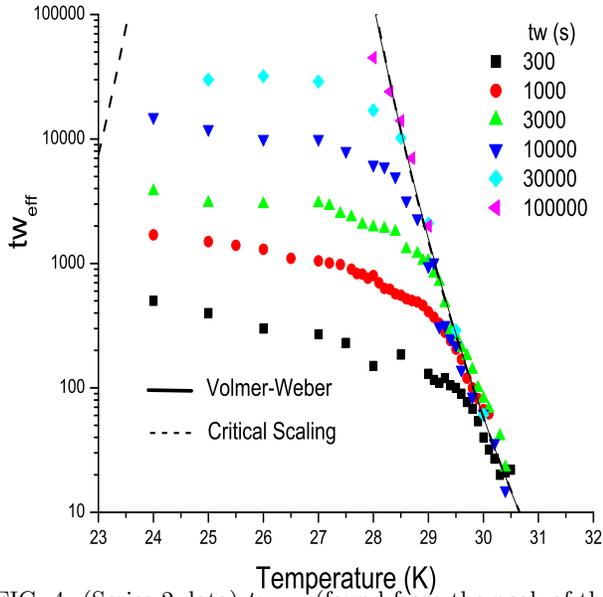}}
 
\caption[]{(Series 2 data) $tw_{eff}$ (found from the peak of the S(t) function), versus temperature for waiting times ranging from 300 s to 100,000 s .}
\end{figure}

\bigskip
\maketitle
\noindent{\bf III.  Discussion}\\

Using $tw_{eff}$ as the governing time scale within the TRM correlation function ($\xi_{TRM}(tw_{eff})$), the power law as a function of temperature can be observed (main graph Fig. 3) over most of the spin glass phase.   The dashed line is a plot of Eq. 2 using tw=1000 s.  Comparing  power law growth of Eq. 2 (tw=1000, dashed line) with $\xi_{TRM}$($tw_{eff}$), we observe that $\xi_{TRM}$ begins to grow more slowly than the straight power law as low as 25K ($\approx 0.8T_g$).  This suppression of the correlated growth continues until the region where $tw_{eff}$ rapidly decreases. In this temperature region the growth of $\xi_{TRM}$($tw_{eff}$ maximizes then decreases as temperature increases. (The same behavior, on a finer temperature grid, is observed in Fig. 5 for Series 2 data)

Figure 4 displays $tw_{eff}$ vs temperature for Series 2 data, for waiting times ranging from 300 s to 100,000 s.  In the region of the peaks observed in Figure 3 (the high temperature region), $tw_{eff}$ overlaps for different waiting times tw, producing a cutoff timescale.  For example at 29K the cutoff timescale is approximately 1000 s. For waiting times less than 1000 s, $tw_{eff}$ still varies with tw.  TRM measurements with waiting times greater than 1000 s all produce a $tw_{eff}$ = 1000 s.  Therefore, in the region of overlap, the time scale for growth of spatial correlations is also cutoff. 
\begin{figure}[t] 
\centering
\resizebox{8cm}{10cm}{\includegraphics{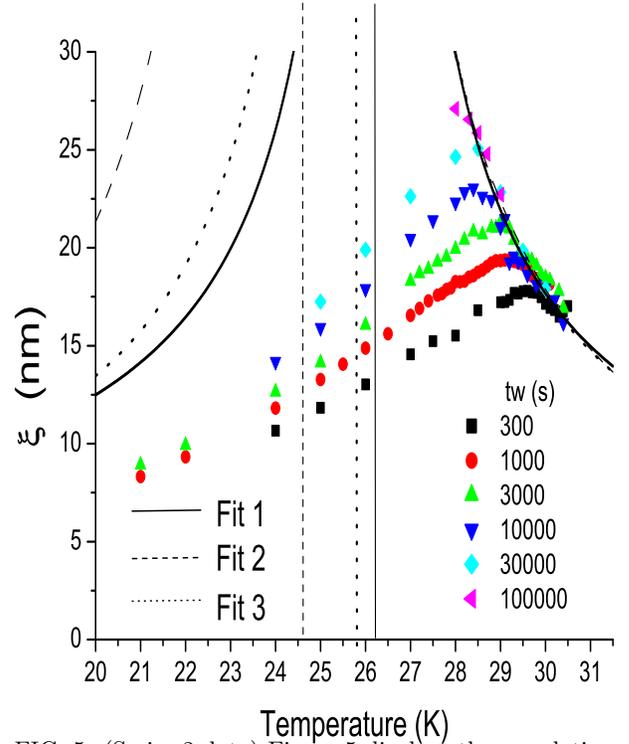}}
\caption[]{(Series 2 data) Figure 5 displays the correlation length found by inserting the values of $tw_{eff}$ displayed in Figure 4 as the timescale for growth of correlations.} 
\end{figure}

There have been many observations of time dependencies in spin glasses near Tg. These include a large number of ac susceptibility measurements, as well as  changes in the  characteristic time scale, as a function of temperature,  observed by Uemura et al. \cite{Uemura} in AuFe and CuMn spin glasses using muon-spin relaxation and  Mezei et al.\cite{Mezei} using neutron spin-echo decay. These experiments all show rapid time scale changes for short timescales in the temperature region above Tg. Fig. 4 is the first observation of rapid changes in the characteristic timescale of the TRM and occurs in the temperature region below Tg. In the past, these time dependencies have been fit to both a Volger-Fulcher Law, indicating a glass transition, and to a power law, indicative of a phase transition. In both cases, glasses or a phase transition,  time scales generally slow down  as the glass or critical temperature is approached.

In glasses the relaxation timescales $\tau$ are proportional to the viscocity $\nu$. As the viscosity increases near the glass temperature the timescales increase as:
\begin{equation}
\large{
\tau=\tau_0{e^{\left(A \over {T-T_o}\right)}} ~~,}
\end{equation}
Setting $\tau=tw_{eff}$ we explored fitting of Eq 4 to the data in Fig. 4. For strong glasses $T_o = 0 K$, we  do not find any fitting parameter that would support this low of a transition temperature. The solid line Eq. 4, in Fig.4 fits the data with $\tau_0 = 2x10^{-s}$ s, A= 201 K and $T_0 = 21.7 K$.

While it is interesting that the time dependence of the data can be fit to glassy dynamics, it is the general consensus that the spin glass exhibits a phase transition.  We fit the time scales $tw_{eff}$ (in the region of overlap, Fig. 4 dashed line) to the dynamic scaling function
\begin{equation}
tw_{eff}=\tau_o{\left|{{T-T_c} \over T_c} \right|}^{-z\nu}~~,
\end{equation}
we obtain best fit values of $\tau_0=2x10^{-8}$ s  for  $T_c=25.8~K$ and $z\nu = 12$.

The fits in Fig. 4 show that the data can be fit to both a glass and phase transition. These transitions have different physics, with the glass transition determined by the rapidly increasing relaxation time scales whereas a phase transition is accompanied by the growth of correlations with $ \xi \longrightarrow \infty$ as $T_c$ is approached.

Fig. 5 is a plot of  $\xi_{TRM}$($tw_{eff}$) vs Temperature for the  Series 2 data where $\xi_{TRM}$($tw_{eff})$ is determined from Eq. 2.   At a given measuring temperature (e.g. 29K), the waiting effect persists for small waiting times (tw $ <$ 1000 s) albeit with reduced $tw_{eff}$ ($tw_{eff}$ $ < $ tw). As with the case of the thin film analysis, it is assumed that $\xi_{TRM}$ grows isotropically until it is confined by a limiting length scale. In the single crystal sample isotropic growth is expected. At 29K the waiting time effect disappears for tw $ > $ 1000 s, as all of this data share a common $tw_{eff}$. From a spatial point of view, at 29 K, $\xi_{TRM}$ can grow up to a finite size  of $\approx$ 19 nm and then ceases to grow.  The line of data  that signals the end of growth of $\xi_{TRM}$ is temperature dependent and for $tw  > tw_{eff}$, time independent. Following the assumption made in the confining geometry of the spin glass films, we conjecture that $\xi_{TRM}$($tw_{eff})$, in the high temperature region, is confined by the size of critical correlations $\xi_{critical}(T)$. This is suggested by the effect on the time dependence, the proximity of the line formed by the length scales  to $T_g$, and  the structure of the line formed by this data.

In Critical Phase Transition Theory,\cite{Fisher83} phase transitions are characterized by the growth of critical correlations and  the growth of a correlation length scale according to the form. 
\begin{equation}
\xi_{crit}(T)=A{\left|{{T-T_c} \over T_c} \right|}^{-\nu}~~,
\end{equation}
It is only at $T_c$ that $\xi_{crit}\rightarrow\infty$. In a real experimental samples, $\xi_{crit}$ will be cut off by finite size effects. Above $T_c$ correlations are limited, but according to Eq. 4 the mean length scale is only a function of temperature.

We fit the high temperature correlation length scales (temperatures greater than the peaks observed in Fig. 5, 40 points)),  to the functional form for $\xi_{crit}$ using both manual and automated search routines to find the minimum in $\chi^2$ within the three dimensional space defined by A, $T_c$ and $\nu$.  We find that there are two fits which minimize and give the same value of $\chi^2$. The solid lines plotted as Fit 1 in Fig. 5 are the fit of Eq. 4 with $T_c=26.2~K$ (.83$T_g$), $\nu= .7$ and minimum spin glass size of $A=4.5~nm$. Fit 2, the dashed lines, show best fit values of $T_c = 24.6~K$ (0.78$T_g$), $\nu = 1.1$,  and $A=3.4~nm$. 

Interestingly the transition temperature found from fitting the effective times in Fig. 4 falls between the two transition temperatures found by fitting the correlation length. Setting  the value of $T_c=25.8 K$ and fitting with that constraint (Fit 3 dotted line) we obtain values of $\nu= .83$ and a minimum spin glass size of $A=3.9~nm$ Using this value of $\nu$, a value for the dynamic exponent of  z=14.5 can be extracted.

Numerical studies indicate a phase transition in spin glasses and this is usually determined from the temperature ($T_c$) at which the Binder cumulant, for different size systems, exhibits crossover. Numerical studies on 3D Ising spinglasses by Fernandaz et al.\cite{Fernandez2015} indicate that Eq. 2 holds for $T<T_c$ but the exponent $c_2T/T_g$ becomes a constant $Z\approx1/6$ above $T_c$. Fitting our data above 26 K to this growth function, we again observe a reduction in $\xi_{TRM}$ above 28 K. Fitting this cutoff length scale to Eq. 4 and using the $\nu= 2.56$ from the same paper, we find $T_c=24.2~K$ and $A=2.2~nm$. We find however that the  correlation length scales are an order of magnitude greater than those shown in Fig. 5. To compare with the numerical values\cite{Fernandez2015} using $\nu= 2.56$, and $T_c=24.2~K$ from above, we find $z=4.7$and $\tau=1.5x10^{-8}~s$.

 \begin{figure}[b] 
\resizebox{6cm}{4cm}{\includegraphics{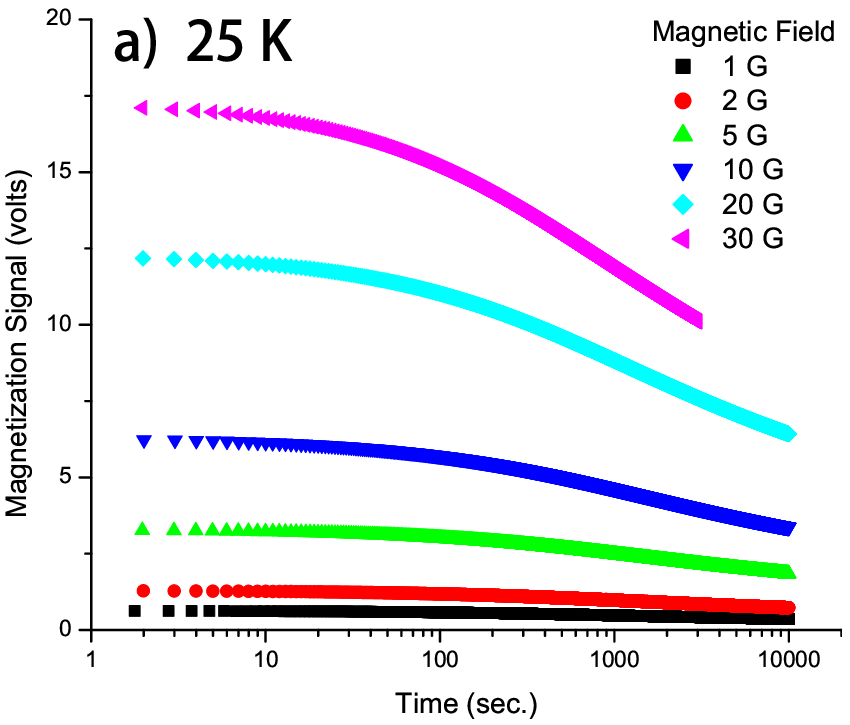}}
\resizebox{6cm}{4cm}{\includegraphics{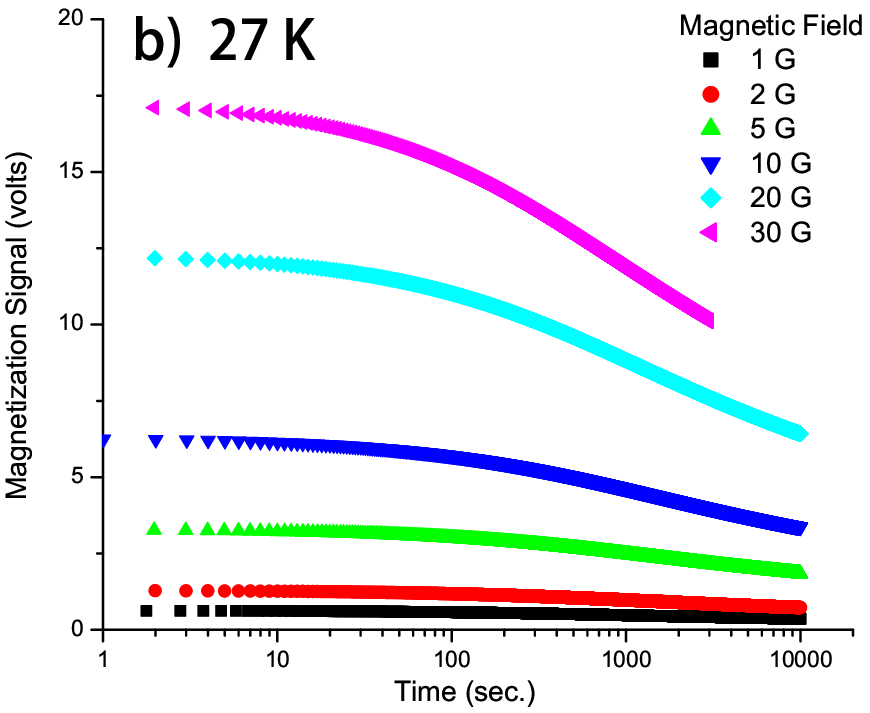}}
\resizebox{6cm}{4cm}{\includegraphics{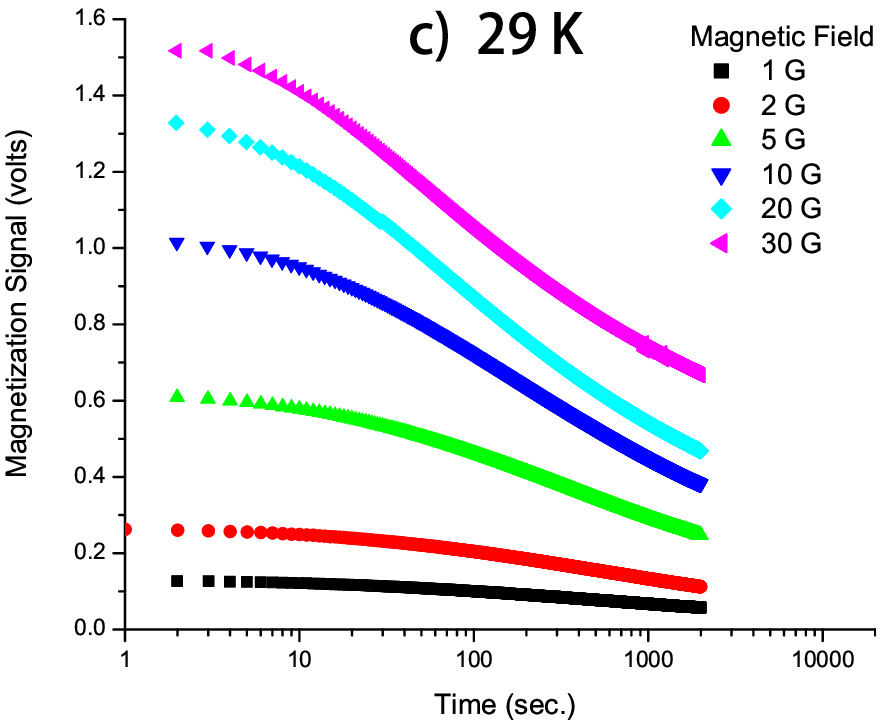}}
\resizebox{6cm}{4cm}{\includegraphics{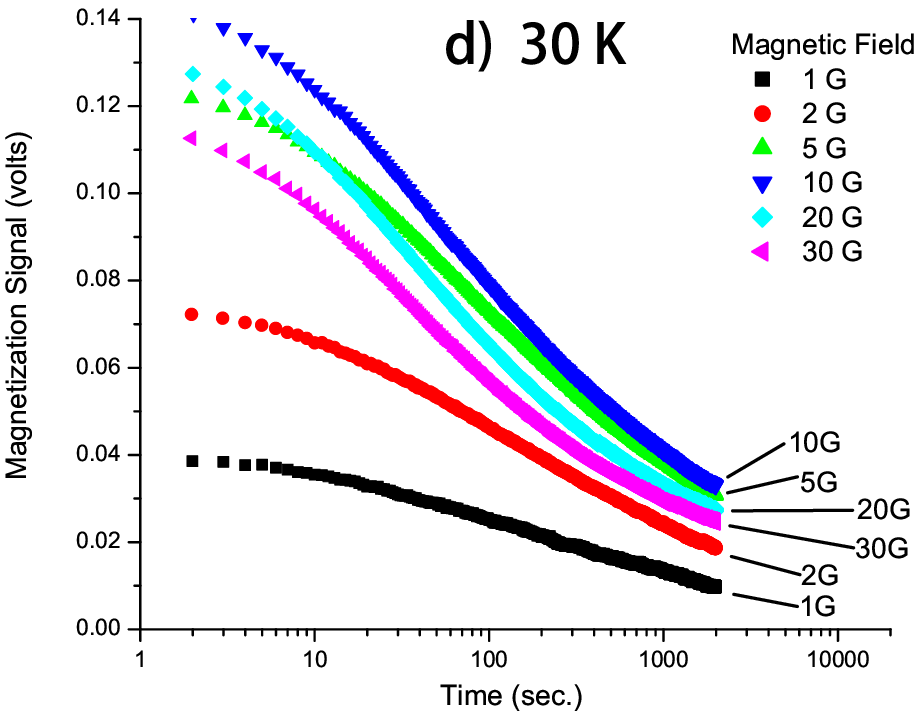}}
\smallskip
\captionsetup{format=plain,justification=centerlast} 
\caption{Variation of TRM decays as a function of Magnetic Field at temperatures ranging between 25 K and 30 K for a 1,000 s waiting time. Magnetic fields of 1, 2, 5, 10, 20 and 30 G were measured.}
\end{figure}

\begin{figure}[t] 
\resizebox{6cm}{4cm}{\includegraphics{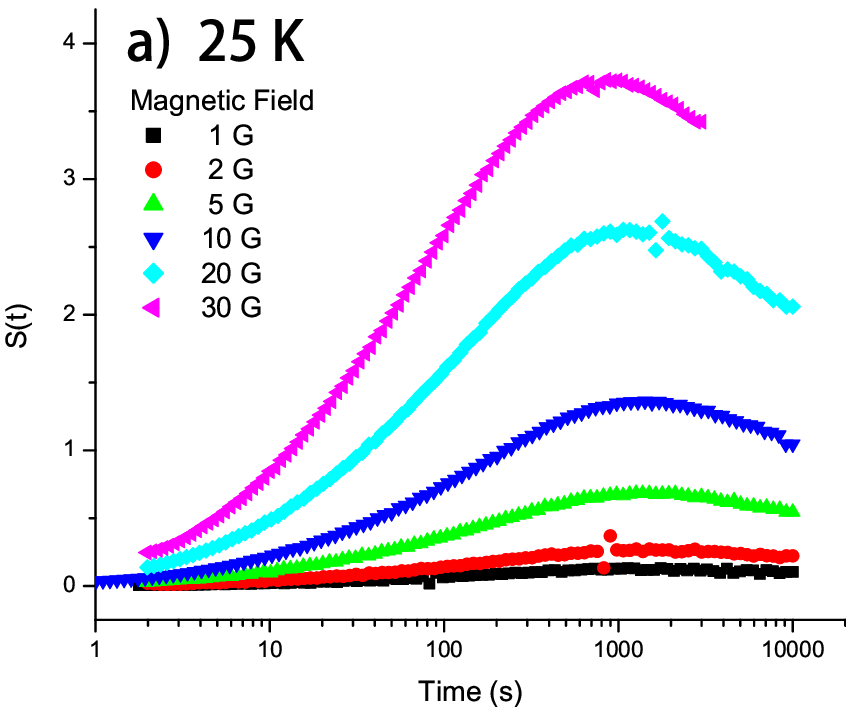}}
\resizebox{6cm}{4cm}{\includegraphics{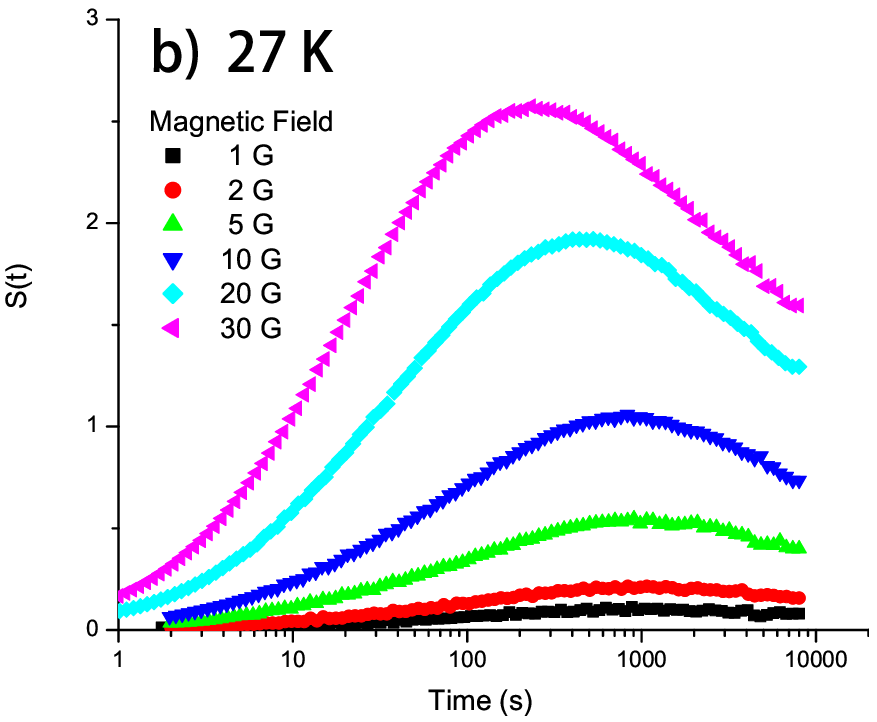}}
\resizebox{6cm}{4cm}{\includegraphics{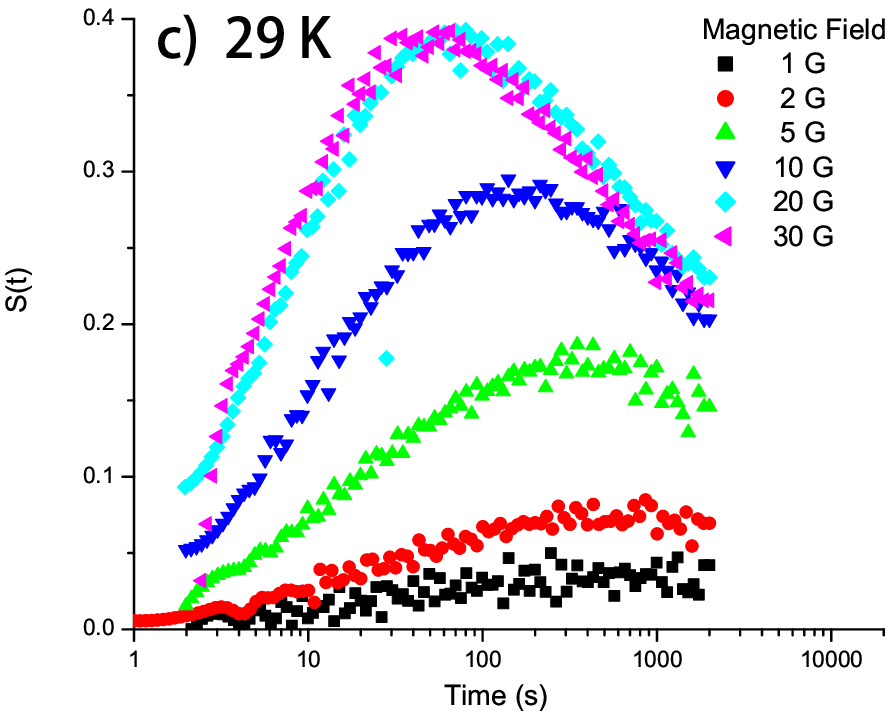}}
\resizebox{6cm}{4cm}{\includegraphics{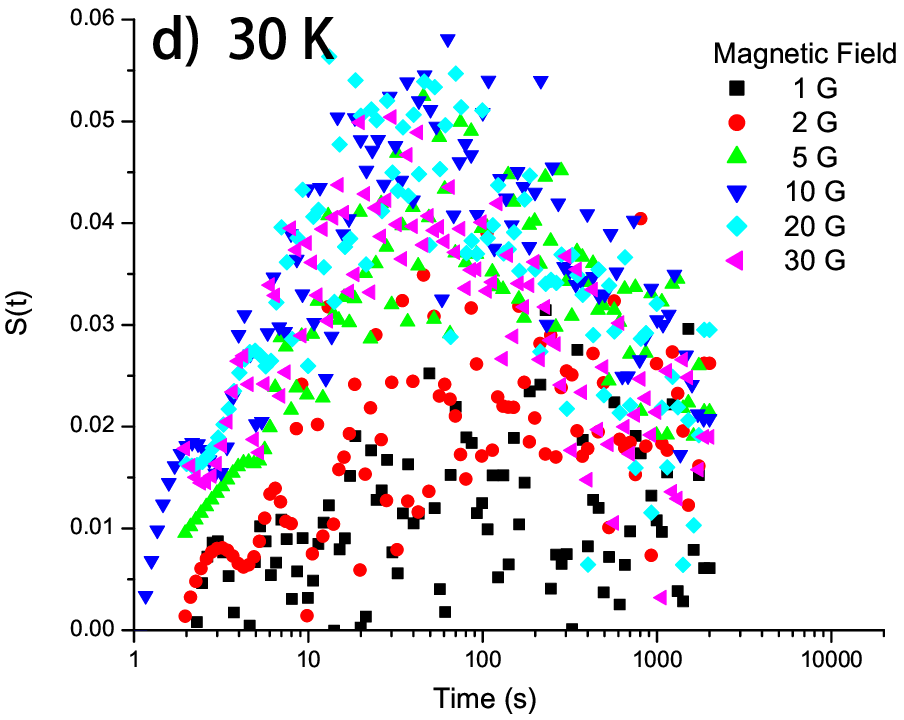}}
\smallskip
\captionsetup{format=plain,justification=centerlast} 
\caption{ Variation of S(t) function (data in Fig. 6) as a function of Magnetic Field at temperatures ranging between 25 K and 30 K for a 1,000 s waiting time. Magnetic fields of 1, 2, 5, 10, 20 and 30 G were measured.Figure 1b displays the  of the same data. The inset in figure 1b shows the three highest temperature data sets on an expanded scale.}
\end{figure}
\begin{figure}[t] 
\centering
\resizebox{6cm}{6cm}{\includegraphics{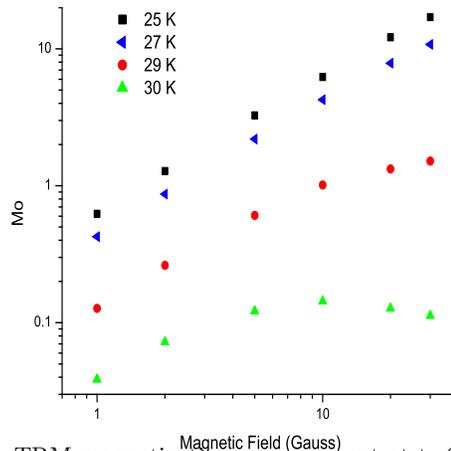}}
\caption[]{TRM magnetization measurement at t=0 s (solid symbols) for the data in Fig. 6. } 
\end{figure}


\bigskip
{\bf IV.  Magnetic Field Dependence near Tg}
\bigskip

We have followed up the above results with a set of field dependent measurements in the "critical region" (25-30K). Figures 6 and 7 display the field dependence of the TRM decay in fields ranging from 1 G to 30 G. It can be observed that at 25 K and 27 K the data looks approximately linear in magnetic field. At 29 K the data become non-linear at high fields and at 30 K the magnetization hits a maximum at 10 G and then decreases as the field gets larger.

In Fig. 7 it can be observed that as the field increases there is a decrease in $tw_{eff}$ (the peak time of the S(t) function). This same effect was observed by Zhai et al.\cite{zhai20} in a generally equivalent measurement to the TRM, the ZFC magnetization decay. We see that the effect is less pronounced at the lowest temperature (25 K), increases through 27K and maximizes at 29 and 30K. This is in the critical region observed in Fig. 4.  We believe that this effect is due to the variation of $T_g$ with magnetic field. In 1993, Kenning et al. mapped the decrease in $T_g$ with increasing magnetic field in a CuMn(6$\%$) polycrystalline sample. This was interpreted as evidence for the AT line, theoretically found as a line of instability of the free energy of the Sherrington Kirkpatrick model, predicting that $T_g$ decreased as $H^{2/3}$. We believe that the reduction of $tw_{eff}$ as a function of H in Fig. 7 is due to a reduction of $T_g(H)$, as a function of H. At a particular temperature T, as the field increases $T_g(H)$ decreases thereby increasing the reduced temperature $T_r = T/T_g(H)$. Therefore at a fixed temperature the decays measured at larger magnetic fields are equivalent to measuring at a higher temperature. Fig. 4 shows that as the temperature increases  $tw_{eff}$ decreases.

In Fig. 8, the initial point in the decay $M_o$ as a function of H is plotted. The plot is log-log graph to express all of the data. Again, the data at the lowest temperature appears to vary approximately linearly suggesting we are far away from the transition temperature. The data approaching $T_g$ looks highly nonlinear implying $T_g$ may be close to the transition temperature. Again we face the issue that Tg is a function of field. We propose to measure the non-linear terms while compensating for the shift in $T_g(H)$.
We propose to first use the Quantum Design magnetometers (SQUIDs) to obtain accurate FC/ZFC curves for 1, 2, 5, 10, 20 and 30G magnetic fields. We will plot the difference between these curves to determine the onset of irreversibility, defining $T_g(H)$.  We will then perform TRM experiments over temperatures ranging from 0.8Tg(H) to Tg(H) as a function of magnetic field. We believe that compensating for the shift in $T_g(H)$ as a function of H is crucial for a correct scaling analysis. Over the range of fields we will probe, the shift in $T_g(H)$ can be as much as 1 K.  In Fig. 4, in the critical region, variations of 1 K can change both $tw_{eff}$ (Fig. 4) and the magnetization (Fig. 2) by as much as an order of magnitude or more completely changing the relationships observed in Fig. 8.

In Summary, we have made extensive measurements of the waiting time effect near the transition temperature $T_g$. For temperatures less than but approaching $T_g$, we observe both the collapse of the observed remanence and the timescale $tw_{eff}$ associated with the waiting time effect. Determining the length scale associated with the growth of correlations in the spin glass phase, we observe, near $T_g$, a cutoff length scale. We associate this cutoff length scale with the critical correlation length scale and determined values for scaling parameters.

\bigskip
{\bf V.  Acknowledgements}
We would like to thank R.L. Orbach, D. Tennant, Q Zhai, V.M. Mayor and E.D. Dahlberg for useful discussions. This work was supported by the U.S. Department of Energy, Office of Science, Basic Energy Sciences, under Award DE-SC0013599. The IUP Dual DC SQUID magnetometer was built under an NSF MRI, Award No. 0852643.  Single crystal growth was performed at the Ames Laboratory which is supported by the Office of Science, Basic Energy Sciences, Materials Sciences and Engineering Division of the U.S. Department of Energy (USDOE), under Contract No. DE-AC02-07CH11358.

\end{document}